\documentclass{bmvc2k}

\title{HypervolGAN: An efficient approach for GAN with multi-objective training function}

\addauthor{Jingwen Su}{jingwen.su@manchester.ac.uk}{1}
\addauthor{Hujun Yin}{hujun.yin@manchester.ac.uk}{1}

\addinstitution{
 Department of Electrical and Electronic Engineering\\
 School of Engineering
 University of Manchester\\
 UK
}

\runninghead{Jingwen Su, Hujun Yin}{HypervolGAN}
\begin{document}

\maketitle

\begin{abstract}
Since the advent of generative adversarial networks (GANs), various loss functions have been developed and combined to constitute the overall training objective function, in order to improve model performance or for specific learning tasks. For instance, in image enhancement or restoration, there are often several criteria to consider such as signal-noise ratio, smoothness, structures and details. However, when the optimization goal has more than one adversarial loss, balancing multiple losses in the overall function becomes a challenging, critical and time-consuming problem. In this paper, we propose to tackle the problem by means of efficient multi-objective optimization. The proposed HypervolGAN adopts an adapted version of hypervolume maximization method to effectively define the multi-objective training function for GAN. We tested our proposed method on solving single image super-resolution problem. Experiments show that the proposed HypervolGAN is efficient in saving computational time and efforts for fine-tuning weights of various losses, and can generate enhanced samples that have better quality than results given by baseline GANs. The work explores the integration of adversarial learning and optimization techniques, which can benefit not only image processing but also a wide range of applications.  
\end{abstract}

%-------------------------------------------------------------------------
\section{Introduction}
Generative adversarial networks (GANs) have drawn a great deal of attentions recently as a powerful framework to generate high perceptual quality images.
Performance superiority of GANs as an alternative to train generative models has been demonstrated by many applications in a variety of areas including image-to-image translation \cite{zhu2017unpaired}, image inpainting \cite{yeh2016semantic}, style transfer \cite{li2016combining}, image restoration \cite{yu2018underwater}, and image synthesis \cite{zhang2017stackgan}.\par
In order to enhance the quality of generated images, various losses have been proposed and combined with the adversarial loss to form the overall training objective function of the GANs. With multiple losses or constraints, balancing between different losses becomes a critical issue for model performance optimization. A linear combination of losses does not guarantee optimal solutions when the objective space is non-convex. Besides, in most GAN models, weightings of various losses are defined empirically with limited explanation on how to derive the best value of weights. Fine-tuning these parameters can be very time-consuming and wasteful of computation resources. We address this problem by considering the combination of losses in the training objective function of GAN as a multi-objective optimization problem. We propose an adapted formulation based on hypervolume indicator to  define the multi-objective function flexibly and efficiently. The resulting GAN is termed as HypervolGAN. \par
% To validate its effectiveness, experiments are performed on SRGAN \cite{ledig2017photo} and Enhanced SRGAN (ESRGAN) \cite{wang2018esrgan}. Our results show that hypervolume maximization provides an efficient and effective approach for finding optimum solution for GAN model and generated images have at least comparable quality as these given by the baseline model.
The remainder of this paper is structured as follows. Section 2 reviews image super-resolution, GAN and multi-objective optimization. Section 3 explains the proposed HypervolGAN in detail. Section 4 presents experiment settings, results and discussions. Finally, Section 5 concludes the work and discusses possible directions of further research.

\section{Related Work}
\subsection{Image Super-resolution}
Super-resolution (SR) algorithms restore a high resolution (HR) image from one or multiple low resolution (LR) observations and have recently become an active topic of research topic due to potentials in a number of practical and real-world applications, such as ultrasound imaging \cite{errico2015ultrafast}, aerial imaging \cite{akgun2005super}, video enhancement \cite{bishop2003super}, and digital holography \cite{fournier2017pixel}. The aim is to provide fine texture details that are absent due to limited capability of the imaging devices in capturing more pixels per unit area of the sensors. Based on the number of LR observations, SR algorithms can be categorized into single image SR (SISR) and multiple images SR (MISR), and our focus here is the SISR problem.\par
Conventional SISR algorithms are based on reconstruction methods by utilizing image priors. Domain-based SISR algorithms use specific class of image priors \cite{tappen2012bayesian,sun2012super}, while generic SISR algorithms use general image priors like edges \cite{fattal2007image}, image statistics (e.g. heavy-tailed gradient distribution in \cite{shan2008fast}), patches \cite{damera2000image} and prediction models that generate HR images through predefined mathematical formula \cite{yang2014single}. In the last decade or so, learning based convolutional neural networks (CNNs) have become the most popular method for SISR and have boosted performances in both accuracy and speed \cite{dong2016accelerating,wang2019end,dong2015image,kim2016accurate,ulyanov2018deep}. As CNNs are becoming to have deeper structures and more complex cost functions, SISR can be considered as an image generation problem and can be solved by GAN models. The most representative studies include SRGAN \cite{ledig2017photo} and ESRGAN \cite{wang2018esrgan}. Also there have been many efforts to develop variants of GAN models with different losses to enhance quality of generated HR images  \cite{dosovitskiy2016generating,chen2017face,yu2016ultra,johnson2016perceptual}.

\subsection{Generative Adversarial Networks}
Introduced in \cite{goodfellow2014generative}, GAN is composed of a generator $G$ and a discriminator $D$, where the generator is a generative model and the discriminator is a classifier network that provides useful gradients for optimizing the generator by adversarial learning, such that generated samples can have higher quality. And the basic objective function of GAN is a min-max game written as,
\begin{equation}
\label{eq:eq1}
\min_{G} \max_{D} {V(D,G)}= \mathbf{E}_{x \sim p_{data}} [\log(D(x))] +\mathbf{E}_{z \sim p_{z}} [\log(1-D(G(z)))] 
\end{equation}
where $x$ represents data sampled from the data distribution $p_{data}$, $z$ represents noise variable sampled from noise distribution $p_{z}$, $D(x)$ represents the probability that input data is from $p_{data}$. Training $G$ and $D$ is simultaneous, the generator is optimized to generate fake samples with plausible details to fool the discriminator, while the goal of the adversarial discriminator is to distinguish fake samples from true samples.\par

Many variants of GAN have been developed as further investigations on adversarial loss and tackling divergence and mode collapse issues in training. Conditional GAN \cite{mirza2014conditional} includes auxiliary information to provide specific data mode. Least squared GAN \cite{mao2017least} adopts the least square function to replace sigmoid cross entropy loss function in the overall training objective function. Wasserstein GAN (WGAN) \cite{arjovsky2017wasserstein} has been proposed to minimize the Wasserstein distance between data distribution and generator distribution. Furthermore, Wasserstein GAN with gradient penalty (WGAN-GP) \cite{gulrajani2017improved} adds a gradient penalty in the overall loss fucntion as an improved version of WGAN. Relativistic GAN (RGAN) and Relativistic average GAN (RaGAN) have been proposed in \cite{jolicoeur2018relativistic} to use a relativistic discriminator, which estimates the probability that true data is more realistic than generated fake data.

\subsection{Multi-Objective Optimization}
Multiple objective optimization (MOO) is a common problem that exists in almost every aspect of the real world, where a compromising and practical solution needs to be derived under the constraints of conflicting objectives. It is also a highly active research topic in  optimization techniques that require objective functions to be optimized simultaneously. A multi-objective optimization problem has $n$ objective functions $f(x)=( f_1(x),...,f_n(x))$ that map a solution $x \in X$ in the decision variable space $X$ to a $n$-dimensional vector $\mathbf{y}=f(x)=(y_1,...,y_n)$ in the objective space $Y$. \par
However, for complex MOO problems, no single solution is capable of realizing simultaneous optimization of several conflicting objectives. Instead, there exists a set of best possible compromising solutions that are called Pareto-optimal solutions. Pareto-optimal solutions are the solutions for which one objective cannot be improved without degrading the others. Without loss of generality, we assume maximization of a MOO problem and there are two decision variables $x_1, x_2 \in X$. $x_1$ is said to dominate $x_2$ (denoted as $x_1 \succ x_2$) if \cite{zitzler1999multiobjective}
\begin{equation}
\begin{aligned}
    &\forall \, i \in \{1,...,n\}: f_i(x_1) \geq f_i(x_2) \qquad and\\
    &\exists \, j \in \{1,...,n\}: f_j(x_1) > f_j(x_2)
\end{aligned}
\end{equation}
A decision variable $x$ is called a nondominated solution if it is dominated by no other variables in the set. For the entire search space, it is a Pareto-optimal solution of the Pareto-optimal set. Corresponding objective vectors are represented by points in the $n$-dimensional objective space that form the Pareto front.\par

There are many multi-objective evolutionary algorithms to derive approximations to the Pareto-optimal solutions for MOO problem \cite{deb2001multi,coello2007evolutionary}. In order to evaluate the quality of the solutions generated by different optimizers, quality indicators have been proposed to map a set of solutions to a scalar value \cite{beume2009complexity,zitzler2003performance}. The hypervolume indicator \cite{zitzler2003performance} is the most useful and representative quality indicator with many favorable properties for performance assessment of multi-objective optimizers \cite{zitzler2007hypervolume}. According to \cite{zitzler1999multiobjective}, the hypervolume indicator $\mathcal{H}$ measures the volume of dominated space bounded by an approximation of Pareto set and the reference point in the $n$-dimensional objective space. One of the ways to calculate $\mathcal{H}$ of a set of solutions $A$ is given by \cite{zitzler2007hypervolume} as follows,
\begin{equation}
    \mathcal{H}(A, r) := \int_r 1[\exists x \in A: f(x) \succeq z \succeq r] \, dz
\end{equation}
where $z \in Y$ represents objective vectors that $A \succeq \{z\}$, $r$ is the reference point that is dominated by all $x$, and $1[\cdot]$ is the indicator operator referring to the attainment function \cite{zitzler2007hypervolume}. 

Maximizing the hypervolume indicator converts the multi-objective optimization problem into single objective optimization, and encourages an approximation set to move towards the Pareto set, thus solutions in the approximation set have better quality values. There are many studies conducted for hypervolume indicator as performance assessment methods \cite{bringmann2013approximation,knowles2002metrics}, guidance for search algorithms \cite{knowles2003bounded,emmerich2005emo}, and fast computation of hypervolume \cite{while2005heuristics,fonseca2006improved}.

\section{Proposed Method}
There is a trend that the training objective function of GAN contains more than one adversarial loss in order to enforce certain constraints such that generated samples can have certain qualities. A convex combination of losses with regularization terms, which is frequently adopted for most of GANs, might not be an effective way to derive efficient solutions (i.e. generated samples) for GANs with multi-objective training function. We propose to solve the problem from the perspective of multi-objective optimization by maximizing the hypervolume of generated samples. And we adapt the computation of hypervolume into a negative logarithm version of the hypervolume enclosed by the objective vectors (i.e. the losses) and their respective upper bounds. Therefore the proposed HypervolGAN has the overall objective function defined as follows,
\begin{equation}
    \label{eq:hypervol}
    \mathcal{L} = - \sum_k log(\mu_k - \mathcal{L}_k)
\end{equation}
where $\mu_k$ denotes the corresponding upper bound for loss $\mathcal{L}_k$. A normalized HypervolGAN regularizes all objective spaces to the range of [0,1], and the overall training objective function becomes,
\begin{equation}
    \label{eq:hypervol_norm}
    \mathcal{L}_{norm} = - \sum_k log(1 - \frac{\mathcal{L}_k}{\mu_k})
\end{equation}
In our case, we implement the training objective function of the proposed HypervolGAN on ESRGAN and SRGAN, thus $\mathcal{L}_k$ includes adversarial loss $\mathcal{L}_{GAN}$, pixel loss $\mathcal{L}_{pix}$ and perceptual loss $\mathcal{L}_{fea}$. For adversarial loss $\mathcal{L}_{GAN}$, ESRGAN adopts the relativistic GAN \cite{jolicoeur2018relativistic} where $\mathcal{L}_{GAN}$ is written as,
\begin{equation}
    \label{eq:esrgan}
    \mathcal{L}_{GAN} = - \mathbf{E} \, [log(1-D(I_{HR},G(I_{LR}))) + log(D(I_{HR},G(I_{LR})))] 
\end{equation}
while SRGAN adopts the conventional adversarial loss equation as follows,
\begin{equation}
    \label{eq:srgan}
    \mathcal{L}_{GAN} = - \mathbf{E} \, [log(D(G(I_{LR})))]
\end{equation}
Pixel loss (Equation \ref{eq:pixelloss}) is a content loss that computes the difference between generated sample $G(I_{LR})$ and ground truth $I_{HR}$ in either L1 or L2 norm. 
\begin{equation}
    \label{eq:pixelloss}
    \mathcal{L}_{pix}= \mathbf{E} \, [\|G(I_{LR})-I_{HR}\|_p]
\end{equation}
Perceptual loss, denoted by $\mathcal{L}_{fea}$, is initially introduced by \cite{johnson2016perceptual} and then extended in SRGAN \cite{ledig2017photo}, which calculates L1 or L2 distance between feature representations of generated samples and ground truth images. 
\begin{equation}
\label{eq:fealoss}
\mathcal{L}_{fea} = \frac{1}{W_{i,j}H_{i,j}} \sum_{x=1}^{W_{i,j}} \sum_{y=1}^{H_{i,j}}  \| \phi_{i,j}(I_{HR})_{x,y} - \phi_{i,j}(G(I_{LR}))_{x,y} \|_p
\end{equation}
where $\phi_{i,j}$ is the feature map obtained after activation of the j-th convolution before i-th maxpooling layer in the VGG19 network. $W_{i,j}$ and $H_{i,j}$ are the dimensions of feature maps. While in ESRGAN \cite{wang2018esrgan}, authors propose to use feature maps before activation.

\section{Experiments}
\subsection{Datasets}
For training, we used DIV2K train dataset \cite{agustsson2017ntire}, which contains 800 high quality images of 2K resolution and their low resolution counterparts with $\times$4 downscaling factor. Model performance was tested on four benchmark datasets, Set5 \cite{bevilacqua2012low}, Set14 \cite{zeyde2010single}, BSDS100 \cite{martin2001database}, and DIV2K test dataset. During training, batch size was set to 16, each patch of the size 128 $\times$ 128. Training data was augmented with random horizontal flips and 90 degree rotations.

\subsection{Experiment Details}
For a fair comparison, we adopted the network architectures of baseline SRGAN and ESRGAN and the training process as introduced in \cite{wang2018esrgan}. Pre-trained PSNR-oriented model was used for initialising the training for ESRGAN, and pre-trained MSE-based super-resolution ResNet for SRGAN training initialization, in order to avoid undesired local optima for the generator \cite{wang2018esrgan,ledig2017photo}. Training took 400k iterations in total. Learning rate was set as $1 \times 10^{-4}$, and halved at 50k, 100k, 200k, and 300k iterations. For optimization, we used the Adam solver \cite{kingma2014adam} with $\beta_1=0.9$, $\beta_2=0.999$. To implement our proposed HpervolGAN, we defined upper bounds for GAN loss, pixel loss and perceptual loss respectively as $\mu_{ESRGAN}=20$, $\mu_{SRGAN}=200$, $\mu_{pix}=0.1$, $\mu_{fea}=10$. And we also investigated the impact of normalization of different objective spaces.
Training ESRGAN took around 3 days to finish, while SRGAN took around 1 day due to its light network structure.
Networks were implemented using Pytorch framework \cite{paszke2017automatic} on a NVIDIA Titan V GPU.

\subsection{Results and Analysis}
\begin{figure}[ht]
\begin{center}
\begin{subfigure}
    \centering
    \includegraphics[width=0.95\textwidth]{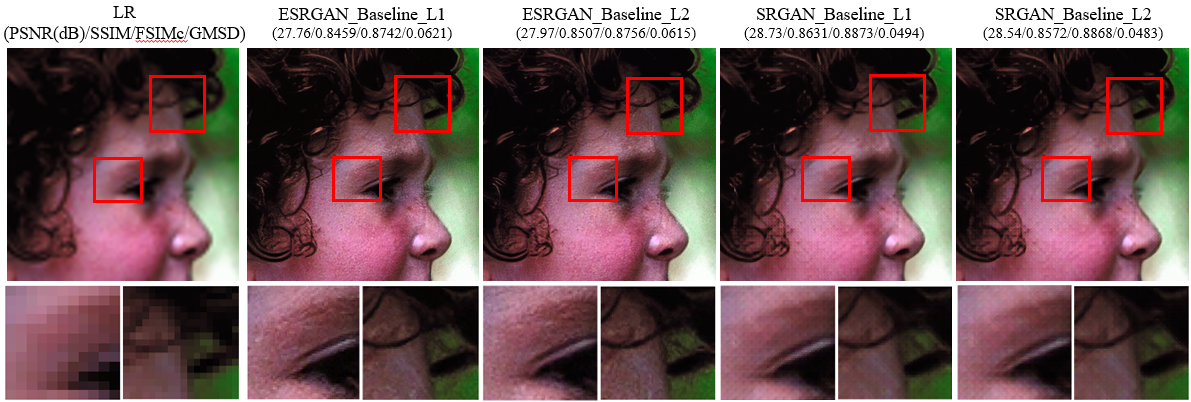}
    % \caption{}
    \label{fig:result1}
\end{subfigure}
\begin{subfigure}
    \centering
     \includegraphics[width=0.95\textwidth]{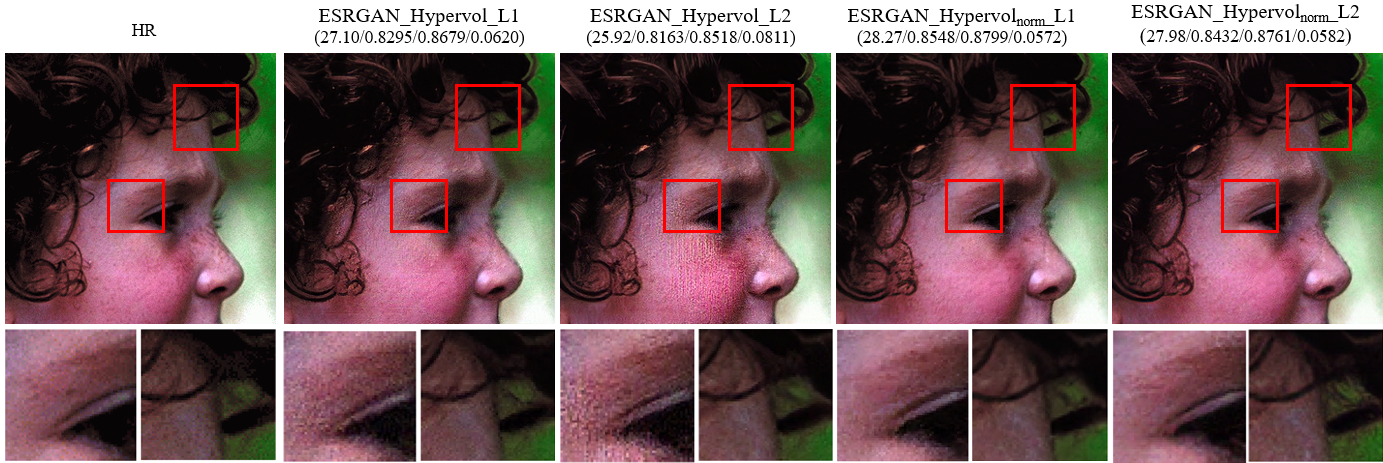}
    % \caption{}
    \label{fig:result2}
\end{subfigure}   
\begin{subfigure}
    \centering
     \includegraphics[width=0.95\textwidth]{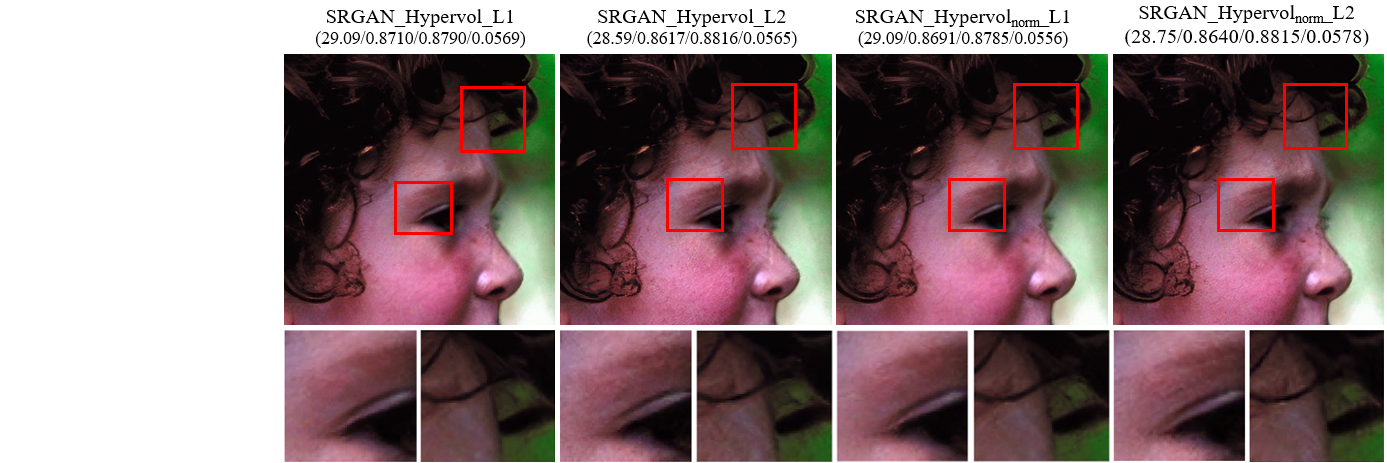}
    % \caption{}
    \label{fig:result3}
\end{subfigure}    
\end{center}
\caption{Exemplar results of Head.PNG from Set5 dataset generated by  trained models with various training objective functions.}
\label{fig:Resultimages}
\end{figure}

\begin{figure}[ht]
\begin{center}
\begin{subfigure}
    \centering
    \includegraphics[width=0.95\textwidth]{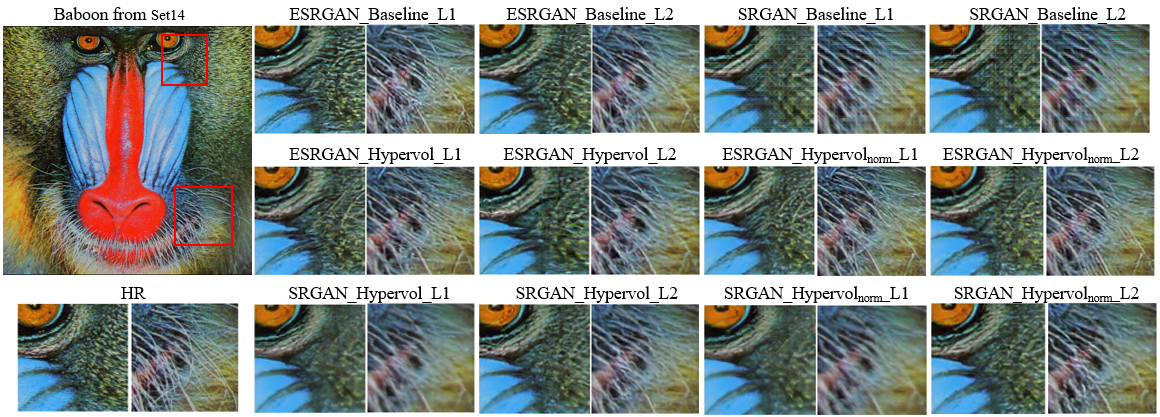}
    % \caption{}
    \label{fig:result4}
\end{subfigure}\\
\begin{subfigure}
    \centering
    \includegraphics[width=0.95\textwidth]{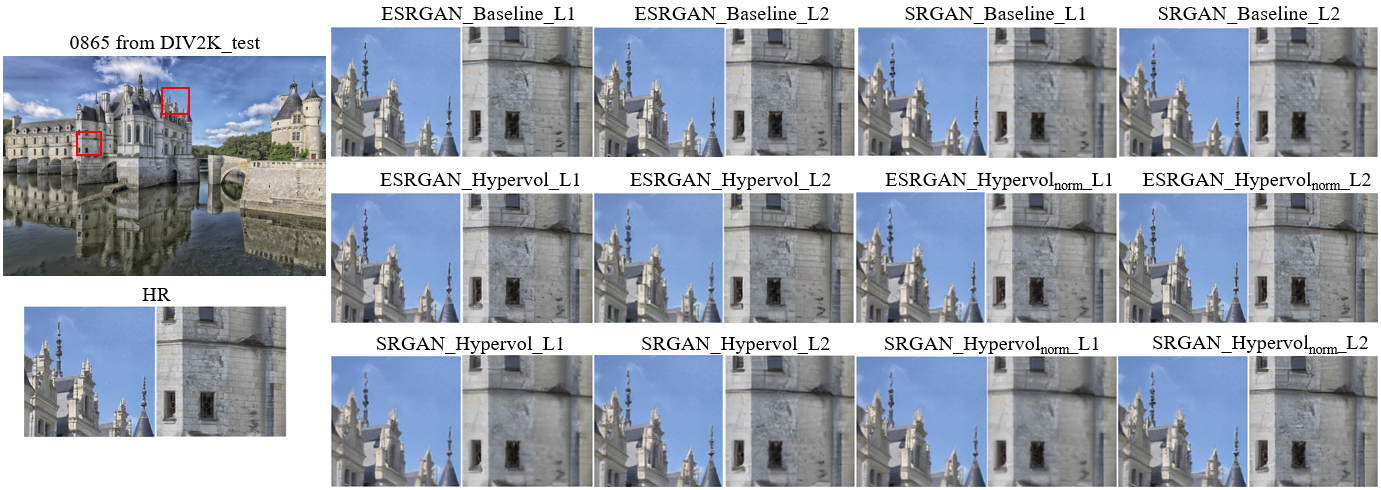}
    % \caption{}
    \label{fig:result5}
\end{subfigure}\\   
\begin{subfigure}
    \centering
     \includegraphics[width=0.95\textwidth]{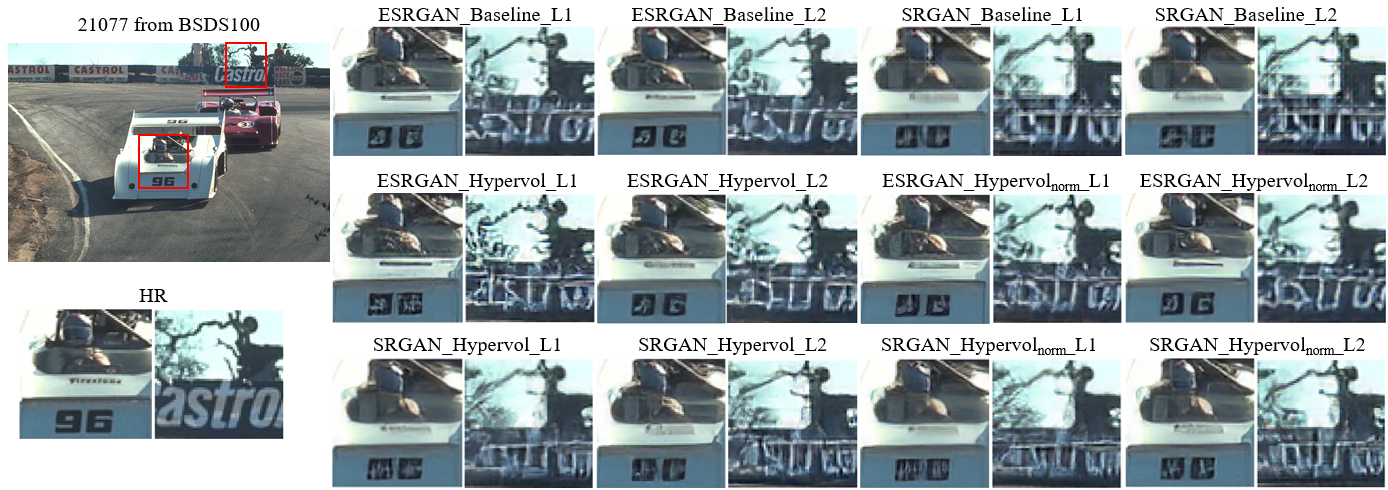}
    % \caption{}
    \label{fig:result6}
\end{subfigure}    
\end{center}
    \caption{Exemplar patches of samples from Set14, DIV2K Test and BSDS 100 datasets generated by trained models with various training objective functions.}
    \label{fig:resultpatches}
\end{figure}

We compared our proposed HypervolGAN with the baseline GAN models on four benchmark datasets. And model performances were evaluated by four representative image quality measures, PSNR, SSIM \cite{wang2004image}, FSIMc \cite{zhang2011fsim} and GMSD \cite{xue2013gradient}. Quantitative results (averaged over three independent runs) are given in Tables \ref{tab:ESRGAN} and \ref{tab:SRGAN}. For visual comparison, exemplar results are provided in Figures \ref{fig:Resultimages} and \ref{fig:resultpatches} with quantitative measures and small patches for inspecting textural details.\par
As can be seen from Tables \ref{tab:ESRGAN} and \ref{tab:SRGAN}, the proposed HypervolGAN outperformed the baseline model in both ESRGAN and SRGAN cases. We believe this is because the training objective function defined in the HypervolGAN (Equation \ref{eq:hypervol}) provides gradient received by the generator $G$ as follows,
\begin{equation}
    \frac{\partial \mathcal{L}}{\partial \theta_G} = \sum_k \frac{1}{\mu_k-l_k} \frac{\partial l_k}{\partial \theta_G}
\end{equation}
Hence the total gradient is a weighted sum of gradients of different losses, and different gradients are weighted in an automatic way instead of fixed as manually defined in the baseline model. Moreover, by this formulation, as the losses vary in every iteration during training, weights of gradients are accordingly adjusted and follow the principle that the larger the loss, the higher importance the corresponding gradient receives. Therefore, it explains the performance given by HypervolGAN is at least comparable to the baseline model, and in most cases, HypervolGAN effectively improves the model performance. On the other hand, although new parameters  (upper bound $\mu_k$ for corresponding loss $l_k$) are introduced, it reduces repetitive and time-consuming work of fine-tuning the weights. And it is easier to find out a loose upper bound for respective loss through trial experiments, considering at the beginning of training when losses are usually high. Compared to the traditional way of defining weights for multiple objectives, HypervolGAN is a more efficient approach to balance the importance of various objectives.  \par
\begin{table}[h]
    \begin{center}
    \setlength{\tabcolsep}{8pt}
    \renewcommand{\arraystretch}{1.1}
    \begin{tabular}{c c c c c c c}
    \hline
     & \multicolumn{2}{c}{Baseline}  & \multicolumn{2}{c}{HypervolGAN} & \multicolumn{2}{c}{HypervolGAN$_{norm}$}   \\
    \cline{2-7}
    Set5 & L1 & L2 & L1 & L2 & L1 & L2  \\
    \hline
    PSNR(dB) & 28.33 & 28.47 & 28.20 & 27.78 & 28.55 & \textbf{28.58}  \\
    SSIM & 0.8018 & 0.8038  & 0.7945 & 0.7881 & 0.8072 & \textbf{0.8114}  \\
    FSIMc & 0.8779 & 0.8795 & 0.8752 & 0.8683 & \textbf{0.8813} & 0.8783  \\
    GMSD & 0.0433 & 0.0426 & 0.0428 & 0.0451 & \textbf{0.0403} & 0.0413  \\
    \hline 
    Set14 &  &  &  &  &  &   \\
    \hline
    PSNR(dB) & 24.72 & 24.87 & 24.63 & 24.35 & \textbf{24.94} & 24.61  \\
    SSIM & 0.6642 & 0.6670  & 0.6586 & 0.6450 & \textbf{0.6701} & 0.6615  \\
    FSIMc & 0.8407 & 0.8412 & 0.8373 & 0.8289 & \textbf{0.8460} & 0.8330  \\
    GMSD & 0.0720 & 0.0691 & 0.0669 & 0.0725 & \textbf{0.0648} & 0.0719  \\
    \hline
    DIV2K Test &  &  &  &  &  &   \\
    \hline
    PSNR(dB) & 26.58 & 26.64 & 26.64 & 26.26 & \textbf{26.99} & 26.56  \\
    SSIM & 0.7413 & 0.7401  & 0.7395 & 0.7290 & \textbf{0.7531} & 0.7394  \\
    FSIMc & 0.9813 & 0.9810 & 0.9834 & 0.9752 & \textbf{0.9859} & 0.9782  \\
    GMSD & 0.0624 & 0.0613 & 0.0594 & 0.0648 & \textbf{0.0565} & 0.0625  \\
    \hline
    BSDS100 &  &  &  &  &  &   \\
    \hline
    PSNR(dB) & 24.08 & 24.16 & 23.79 & 23.91 & \textbf{24.23} & 24.21  \\
    SSIM & 0.6258 & 0.6288  & 0.6145 & 0.6168 & \textbf{0.6318} & 0.6276  \\
    FSIMc & 0.8032 & 0.8019 & 0.7932 & 0.7938 & \textbf{0.8047} & 0.7989  \\
    GMSD & 0.0814 & 0.0796 & 0.0781 & 0.0822 & \textbf{0.0745} & 0.0794  \\
    \hline
    \end{tabular}   
    \end{center}
    \caption{Performances of various training objective functions for ESRGAN on Set5, Set14, DIV2K test and BSDS100 datasets.}
    \label{tab:ESRGAN}
\end{table}
\begin{table}[ht]
    \begin{center}
    \setlength{\tabcolsep}{8pt}
    \renewcommand{\arraystretch}{1.1}
    \begin{tabular}{c c c c c c c}
    \hline
     & \multicolumn{2}{c}{Baseline}  & \multicolumn{2}{c}{HypervolGAN} & \multicolumn{2}{c}{HypervolGAN$_{norm}$}   \\
    \cline{2-7}
    Set5 & L1 & L2 & L1 & L2 & L1 & L2  \\
    \hline
    PSNR(dB) & 28.40 & 27.76 & 29.37 & 28.18 & \textbf{29.39} & 28.21 \\
    SSIM & 0.8136 & 0.7961& \textbf{0.8377} & 0.8116 & 0.8374 & 0.8116 \\
    FSIMc & 0.8763 & 0.8491 & \textbf{0.8938} & 0.8772 & 0.8933 & 0.8773 \\
    GMSD & \textbf{0.0387} & \textbf{0.0387} & 0.0408 & 0.0466 & 0.0406 & 0.0473 \\
    \hline 
    Set14 &  &  &  &  &  &   \\
    \hline
    PSNR(dB) & 25.21 & 24.64 & 25.98 & 25.10 & \textbf{26.02} & 24.93 \\
    SSIM & 0.6846 & 0.6671 & 0.7019 & 0.6754 & \textbf{0.7052} & 0.6635 \\
    FSIMc & 0.8429 & 0.8268 & 0.8548 & 0.8379 & \textbf{0.8556} & 0.8364 \\
    GMSD & \textbf{0.0646} & 0.0656 & 0.0693 & 0.0745 & 0.0686 & 0.0743 \\
    \hline
    DIV2K Test &  &  &  &  &  &   \\
    \hline
    PSNR(dB) & 27.20 & 26.72 & 28.13 & 26.99 & \textbf{28.15} & 26.86 \\
    SSIM & 0.7630 & 0.7496 & \textbf{0.7878} & 0.7536 & 0.7873 & 0.7506 \\
    FSIMc & 0.9845 & 0.9832 & 0.9859 & 0.9801 & \textbf{0.9860} & 0.9800 \\
    GMSD & \textbf{0.0570} & 0.0574 & 0.0598 & 0.0659 & 0.0594 & 0.0666 \\
    \hline
    BSDS100 &  &  &  &  &  &   \\
    \hline
    PSNR(dB) & 24.78 & 24.43 & \textbf{25.45 }& 24.58 & 25.43 & 24.47 \\
    SSIM & 0.6541 & 0.6403 & \textbf{0.6697} & 0.6370 & 0.6685 & 0.6311 \\
    FSIMc & 0.7923 & 0.7634 & \textbf{0.8108} & 0.8008 & \textbf{0.8108} & 0.7992 \\
    GMSD & \textbf{0.0718} & 0.0722 & 0.0787 & 0.0820 & 0.0789 & 0.0827 \\
    \hline
    \end{tabular}   
    \end{center}
    \caption{Performances of various training objective functions for SRGAN on Set5, Set14, DIV2K test and BSDS100 datasets.}
    \label{tab:SRGAN}
\end{table}

In addition, we also investigated the effects of using different GAN models, L1 or L2 norm for pixel loss and perceptual loss, and normalization of objective spaces. The overall performances can be obtained from the quantitative results (Tables \ref{tab:ESRGAN} and \ref{tab:SRGAN}) and qualitative results (Figure \ref{fig:Resultimages}). SRGAN produces better results and higher quality images with smoother details than ESRGAN, and SRGAN also has lighter network structure and requires shorter training time and less computation resources. With regard to the norm, trained ESRGAN models with losses defined in L1 norm have slightly better performances over those defined in L2 norm, while for trained SRGAN models performance is significantly improved by L1 norm. Although visual difference is not obvious to be observed, L1 norm on pixel loss and perceptual loss is beneficial for training GANs for SISR.
Lastly, normalization of objective spaces has different impacts on ESRGAN and SRGAN respectively. Quantitatively, HypervolGAN$_{norm}$ outperforms HypervolGAN by a larger margin on ESRGAN than SRGAN. Qualitatively, HypervolGAN$_{norm}$ generates more smooth details and enhances image quality as shown in Figure \ref{fig:Resultimages}. In short, normalization of multi-objective spaces is a necessary component for adopting the HypervolGAN approach.
More example patches from other datasets are provided in Figure \ref{fig:resultpatches} for further observations of textural details in generated images.

\section{Conclusions}
In this paper, we have proposed HypervolGAN for training multi-objective training functions for GAN and validated its effectiveness on improving model performance for tackling the single image super-resolution task. Networks trained by different training objective functions have been tested on four benchmark datasets for the task and generated extensive experimental results to confirm the superiority of the proposed HypervolGAN over various GANs. With HypervolGAN, multi-objective training for GANs can concentrate efforts on exploring meaningful components in the overall loss function, and it is flexible to experiment without concerning about balancing weights and wasting computation time and resources. This work provides an initial study on combining adversarial learning and multi-objective optimization, there are many potential relevant research topics to further advance the investigation and generalization. For example, upper bound value adaption for different types of GAN loss and additional constraints, applying HypervolGAN for solving other image processing topics or beyond.
\clearpage
\bibliography{egbib}
\end{document}